\documentclass[10pt, preprintnumbers, twocolumn, amsmath, amssymbre, fleqn, nofootinbib]{revtex4}

\usepackage{amsmath}
\usepackage{amsfonts}
\usepackage{amssymb}
\usepackage[colorlinks,linkcolor=blue,citecolor=red]{hyperref}
\usepackage{graphicx}
\usepackage{hyperref}
\usepackage{subfigure}
\usepackage{graphicx}
\usepackage{dcolumn}
\usepackage{bm}
\usepackage{amsthm,amsfonts}
\usepackage{graphicx}
\usepackage{physics}
\usepackage{hyperref}
\usepackage{lscape} 
\usepackage{amsmath}
\usepackage{amssymb}
\usepackage{graphics}
\usepackage{psfrag}
\usepackage{mathtools}
\usepackage{appendix}
\usepackage{comment}
\usepackage{soul}
\graphicspath{{figs/}}

\newcommand{\nc}{\newcommand}
\nc{\ba}{\begin{eqnarray}}
\nc{\ea}{\end{eqnarray}}
\nc{\bfk}{\bf{k} }
\nc{\bfq}{\bf{q} }
\nc{\bfp}{\bf{p} }

\nc{\rc}{\textcolor[rgb]{1.00,0.00,0.00}}
\nc{\bc}{\textcolor[rgb]{0.00,0.07,1.00}}

\begin{document}

\title{Gravitational Bound State Perturbations Inside Black Holes and Isospectrality}

\author{Hassan Firouzjahi$^1$}

\author{Kazem Rezazadeh$^1$}

\author{Masoud Molaei$^2$}

\affiliation{$^1$School of Astronomy, Institute for Research in Fundamental Sciences (IPM), P. O. Box 19395-5746, Tehran, Iran}

\affiliation{$^2$ Department of Physics, Sharif University of Technology, Azadi Avenue, Tehran, Iran}

\begin{abstract}

We study the bound state solutions for the polar perturbations in the interior of the Schwarzschild black hole. It is shown that for a given value of the spherical harmonic index $\ell$, there are a total of $\ell-1$ bound states for polar perturbations. We show both analytically and numerically that the spectrum of $\ell-2$ of these perturbations coincides exactly with the spectrum of axial perturbations. Consequently, the isospectrality between the bound states of axial and polar perturbations in the interior of the black hole is preserved. Furthermore, the additional mode found in the spectrum of polar perturbations is the algebraically special mode, which also furnishes the ground state of polar perturbations. It is shown that the spectrum of the highly excited states is equally spaced, which, in the semi-classical approximation, yields the black hole area quantization $\Delta A = 16 \pi l_{\mathrm{Pl}}^2$.

\end{abstract}


\maketitle


\section{Introduction}

The unperturbed black hole (BH) solutions of the Einstein field equation are mathematically simple and yet elegant \cite{Chandrasekhar:1985kt}. However, the observable physical effects appear at the level of perturbations, such as in the detection of gravitational waves (GWs) emission during the merging of BHs and the ringdown phase as observed in LIGO/Virgo/KAGRA collaborations \cite{LIGOScientific:2016aoc, LIGOScientific:2017vwq, LIGOScientific:2020zkf, LIGOScientific:2025obp}. A key role in BH physics is played by the BH event horizon, which separates the interior region of the BH from its exterior region. In particular, as the interior region is causally disconnected from the exterior region, it is assumed that the dynamical events happening in the interior region are inaccessible to the exterior observers and hence irrelevant for gravitational and cosmological observations. While this notion is practically viable, the interior of the BH is part of the whole manifold, and a consistent treatment of  BH perturbations requires that the perturbations in the interior of the BH be studied as well.

The question of bound state perturbations in the interior of the Schwarzschild BH was originally studied in \cite{Firouzjahi:2018drr} and was revisited more recently in \cite{Steinhauer:2025bbs, Firouzjahi:2025jja}. For other works
concerning the interior of BH, see also \cite{Fiziev:2006tx, Firouzjahi:2024xbo}. The bound states are interpreted as the solutions with an imaginary spectrum in which the perturbations are regular at the center of the BH, while their time-dependent profile falls off exponentially on the event horizon. It is important to note that these perturbations are independent of perturbations in the exterior region and are defined intrinsically in the interior of BHs.
It is demonstrated in \cite{ Firouzjahi:2025jja} that scalar, vector, and axial tensor perturbations all admit bound state solutions. Specifically, it is shown in \cite{ Firouzjahi:2025jja} that for a given value of the angular index $\ell$ and the spin $s$ of the fields, there are a total of $\ell-s$ bound state solutions.
Finally, it is shown in \cite{Steinhauer:2025bbs} that the bound state solutions inside the BHs have the curious property that their total wave function has a non-zero amplitude on the event horizon. Correspondingly, these bound states are named in \cite{Steinhauer:2025bbs} as overdamped quasibound states (OQBSs).

The quasi-normal modes (QNMs) perturbations in the exterior of BHs are vastly studied, for some reviews and an incomplete list of theoretical studies
see \cite{Nollert:1999ji, Kokkotas:1999bd, Berti:2009kk, Konoplya:2011qq, Berti:2025hly, Iyer:1986np, Motl:2003cd, Andersson:2003fh, Cardoso:2003cj, Medved:2003rga, Berti:2005ys,  Konoplya:2019hlu, Hod:2025scz}. The scalar, vector, and axial tensor perturbations are governed by the Regge-Wheeler (RW) equation, while the corresponding equation for the polar tensor perturbations is the Zerilli equation \cite{Regge:1957td, Zerilli:1970se, Chandrasekhar:1985kt}. The RW and Zerilli potentials have different algebraic forms, but there is a one-to-one transformation which links the polar perturbations to the axial perturbations \cite{Chandrasekhar:1985kt, Chandrasekhar:1975zza, Anderson:1991kx}.  Since this transformation is regular in the exterior region, an immediate conclusion is that the axial and polar perturbations have the same spectra of QNMs perturbations.

In this work, we extend upon the analysis of \cite{Steinhauer:2025bbs, Firouzjahi:2025jja} and study the bound state perturbations for the polar perturbations in the interior of the Schwarzschild BH.
It is important to note that the bound state solutions considered here and in \cite{Firouzjahi:2018drr, Steinhauer:2025bbs, Firouzjahi:2025jja} are different from the bound state solutions of the inverted potential proposed by Mashhoon and collaborators \cite{Mashhoon:1982imb, Blome:1981azp, Ferrari:1984zz, Ferrari:1984ozr, Liu:1996cxr}. More specifically, the bound states in \cite{Mashhoon:1982imb, Blome:1981azp, Ferrari:1984zz, Ferrari:1984ozr, Liu:1996cxr} are obtained from the complex transformation of the RW equation to bring the potential in the form of Poschl-Teller or the Eckart potentials, which can be solved analytically. This method was further employed recently in \cite{Volkel:2025lhe} (see also \cite{Hatsuda:2019eoj}) in which the inverted RW potential is considered directly, and a numerical method is used to determine the eigenvalues of energies. \\

\section{Gravitational Perturbations Inside the BH}

The Schwarzschild metric in spherical coordinates is given as usual by,
\ba
\label{metric-Sch}
\dd s^2 = - \big( 1- \frac{1}{r}\big) \dd { t\, }^2 + \frac{\dd r^2}{\big(
	1- \frac{1}{r}\big)} + r^2 \dd \Omega^2 \, ,
\ea
in which $\dd \Omega^2$ represents the angular part of the metric. We have rescaled the dimensionful coordinate by $r \rightarrow 2 G M r$ (with $M$ and $G$ being the BH mass and the Newton constant)
so the event horizon is located at $r=1$.

As usual, let us define the tortoise coordinate $d r_*= \dd r \big( 1- \frac{1}{r}\big)^{-1}$. Note that this differential relation between $r_*$ and $r$ is valid for both interior and exterior regions. Also, since the roles of $r$ and $t$ are switched in the interior of BH, the coordinate $r_*$ plays the role of the time while $t$ is now spacelike. 
We expand the perturbations in terms of the spherical harmonics and denote the remaining parts of the wave function by $Z(r_*, t) \propto e^{-i \omega t} Z(r_*)$.  As in standard BH perturbations theory, the equations governing the axial and polar perturbations take the following form,
\ba
\label{master}
\frac{\dd^2 Z(r_*)}{\dd r_*^2} + (\omega^2 - V_{\mathrm{eff}}) Z(r_*)=0 \, ,
\ea
in which $V_{\mathrm{eff}}$ takes either the RW or the Zerilli potentials denoted by $V^{-}$ and $V^{+}$ respectively. Similar to  \cite{Steinhauer:2025bbs, Firouzjahi:2025jja}, we look for bound state solution 
$\omega= i  \omega_I$ with $\omega_I >0$.

As it is well-known, there are relations between $V^{\pm}$ and between the corresponding mode functions $ Z^\pm$ which ensure the isospectrality of the QNM spectrum associated to the
axial and polar perturbations in the exterior region \cite{Chandrasekhar:1985kt, Chandrasekhar:1975zza}. More specifically,
\ba
\label{trans1}
V^\pm = \pm 3 \frac{\dd f}{\dd {r_*}} + 9 f^2 + 4 \lambda (\lambda+1) f \, ,
\ea
and
\ba
\label{trans2}
\Big(4 \lambda (\lambda+1) {\mp} 6 i \omega \Big) Z^\pm
&=& \Big[ 4 \lambda (\lambda+1) + \frac{18 (r^2- r)}{r^3 ( 2 \lambda r + 3) }
\Big] Z^{\mp} \nonumber\\
&{\pm}& 6 \frac{\dd Z^{\mp}}{ \dd r_*} \, ,
\ea
in which $2\lambda\equiv (\ell-1) (\ell+2) $ and,
\ba
 \hspace{1cm} f\equiv \frac{r- 1}{ r^2 (2\lambda r +3) } \, .
\ea
Indeed, one can specifically check that starting with $Z^{\pm}$ as the solution of the master equation (\ref{master}) and using the transformations (\ref{trans1}) and (\ref{trans2}), then $Z^{\mp}$ satisfies Eq. (\ref{master}) as well.

\begin{figure}[t!]
\begin{center}
\hspace{-0.5cm}	\includegraphics[scale=0.66]{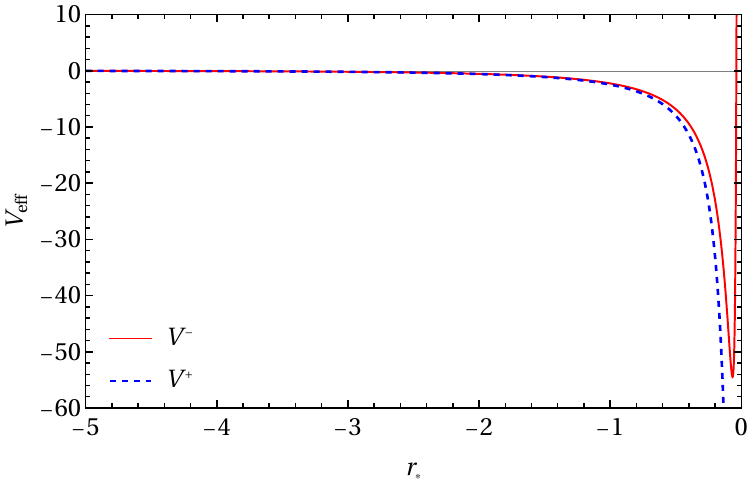}
	\end{center}
\caption{The effective potentials $V^-(r_*)$ and $V^+(r_*)$ given in Eqs. \eqref{V-RW} and \eqref{V-Z} with $\ell = 3$.
\label{fig:Veff}
}
\end{figure}

The specific forms of $V^\pm$ are presented in Appendix \ref{numeric-ap}
while their plots are shown in  Fig. \ref{fig:Veff}. It is seen that $V^-$ develops a negative global minimum before diverging like  $V^- \sim 3/r^4$ near the center.  On the other hand, $V^+$ is always negative and diverges like $V^+ \sim -1/r^4$ at $r=0$.

Intuitively, the transformations (\ref{trans1}) and (\ref{trans2}) may look singular at  $r=0$. More specifically,  the regularity of the bound state solutions at $r=0$ requires that $Z^- \sim r^3$ while $Z^+ \sim r$. Plugging these asymptotic forms in (\ref{trans1}) and (\ref{trans2}) one may conclude that \cite{Steinhauer:2025bbs} $ Z^{-}_{\mathrm{reg}} \rightarrow Z^{+}_{\mathrm{reg}} $ while $
 Z^{+}_{\mathrm{reg}} \rightarrow Z^{-}_{\mathrm{sing}}$. If so, this implies that starting with the regular axial perturbations, one can
obtain the regular polar perturbation, but not the other way around.
However, this is subtle and misleading. We will show that the transformations
(\ref{trans1}) and (\ref{trans2}) are indeed regular and, except for one particular mode,  starting with regular $Z^{\pm}$, one obtains the regular
$Z^{\mp}$. The only exception is the ``algebraically special mode" (ASM) which, for a given value of $\ell$,  turns out to be the ground state of the polar perturbations.

The ASM is the solution in which the left and right hand sides of Eq. (\ref{trans2}) vanish simultaneously. With $\omega_I>0$, the left hand side of Eq. (\ref{trans2}) allows only the lower sign corresponding to
the polar perturbations  with the spectrum \cite{Chandrasekhar:1985kt, Chandrasekhar:1984mgh, Couch:1973zc, Wald:1973wwa},
\ba
\label{special}
\hspace{0.5cm}
\omega=  \omega_\mathrm{sp}\equiv \frac{i}{6} (\ell-1) \ell (\ell+1) (\ell+2)
= \frac{2 i}{3} \lambda (1+ \lambda) \, .
\ea
On the other hand, solving the equation in the right-hand side of (\ref{trans2}),  the profile of the polar perturbations associated with
ASM is obtained to be,
\ba
\label{Z-sp}
Z^+_{\mathrm{sp}}(r) = N_+(\lambda) \frac{r (1- r)^{- i  \omega_\mathrm{sp}}}{3+ 2 \lambda r}
e^{-i \omega_\mathrm{sp} r} \, ,
\ea
in which $N_+(\lambda)$ is a normalization constant.
One can easily check that $Z^+_{\mathrm{sp}}(r)$ satisfies the Zerilli equation (\ref{master}).
On the other hand, since $Z^+_{\mathrm{sp}}(r)$ satisfies Eq. (\ref{trans2}) identically at $\omega= \omega_\mathrm{sp}$, there is no equation left for $Z^-_{\mathrm{sp}}(r)=0$,  meaning that the ASM is not shared by the axial perturbations. This is consistent with the results of \cite{Firouzjahi:2025jja} in which no bound state solutions for axial perturbations at the frequency $\omega=\omega_\mathrm{sp}$ were obtained.

\begin{figure}[t!]
\begin{center}
\hspace{-0.1cm}	\includegraphics[scale=0.66]{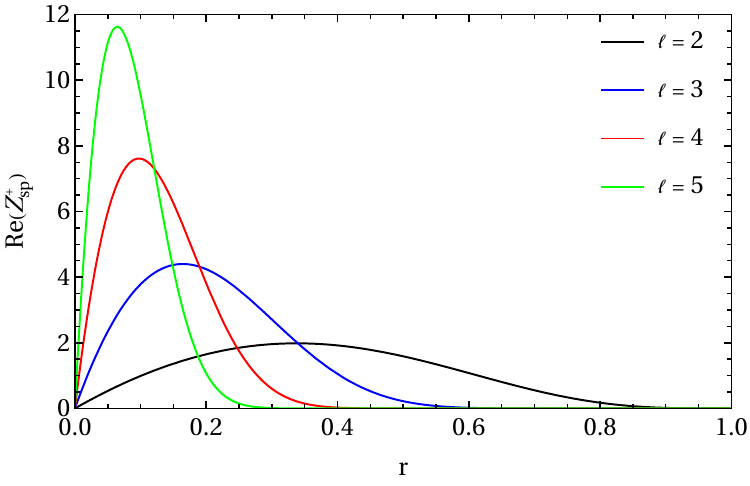}
	\end{center}
\caption{The profile of the normalized algebraically special mode $Z^+_{\mathrm{sp}}(r)$ given in Eq. (\ref{Z-sp}). Curves from bottom to top correspond to $\ell=2, 3, 4, 5$, respectively.
\label{special-fig}
}
\end{figure}

In Fig. \ref{special-fig}, we have presented the plots of $Z^+_{\mathrm{sp}}(r)$ given in Eq. (\ref{Z-sp}) for various values of $\ell$. We have normalized the wavefunction by fixing $N_+(\lambda)$ such that $\int_{-\infty}^{0} \dd r_* | Z^+_{\mathrm{sp}}(r)|^2 =1$. As can be seen, $Z^+_{\mathrm{sp}}(r)$ vanishes at $r=0, 1$ while there is a peak in the intermediate region.

\vspace{0.3cm}

\section{Isospectrality from SUSY Analysis}

Here, using the supersymmetry (SUSY) method,  we analytically confirm the existence of isospectrality between the polar and axial perturbations in the interior of the BH and highlight
the special role which is played by the ASM mode. We comment that this method is formally parallel to the corresponding analysis for the QNMs of the exterior region, see for example  \cite{Cardoso:2001bb, Konoplya:2011qq, Berti:2009kk}. However, there are important differences between the two setups
that require independent investigations. First, we look for the bound state solutions that have different boundary conditions than  QNMs in the exterior region. Second, the potentials $V^{\pm}$ are singular at $r=0$ so the regularity of  transformations (\ref{trans1}) and (\ref{trans2}) is not manifest a priori \cite{Steinhauer:2025bbs}. Here we present the main results, while for
in-depth reviews see \cite{Gangopadhyaya:2017wpf, cooper1995supersymmetry}.


Consider the superpotential,
\ba
\label{SuperPotentoial}
\mathcal{W}_\ell(r)\equiv\frac{3\,(1-r)}{r^{2}(3+2\lambda r)}-\frac{2\lambda(\lambda+1)}{3} = - 3 f+ i \omega_{\mathrm{sp}} \, ,
\ea
in the domain $0\leq r\leq 1$. Equipped with this superpotential,
SUSY ladder operators are constructed as follows,
\ba
\label{SUSYLadder}
    \hat{\mathcal{D}}^{\pm}_{\ell}(r_{*})\equiv \pm\dfrac{d}{d r_{*}}+\mathcal{W}_{\ell}(r_{*}) \, .
\ea
Using the above ladder operators, we construct the partner Schrodinger  Hamiltonians,
\ba\label{3}
 \mathcal{H}_{\ell}^{\pm}(r_*)\equiv \hat{\mathcal{D}}_{\ell}^{\mp}(r_*)\hat{\mathcal{D}}_{\ell}^{\pm}(r_*)=-\dfrac{\dd^2}{\dd r_*^2}
 +U_{\ell}^{\pm}(r_*) \, ,
\ea
with the partner potentials given as,
\ba\label{4}
\hspace{1cm} U_{\ell}^{\pm}(r_*)\equiv\mathcal{W}_{\ell}^{2}(r_*)\mp\dfrac{\dd\mathcal{W}_{\ell}(r_*)}{\dd r_*} \, .
\ea
The  Schrodinger-like equations associated with this system are,
\ba
\label{SUSYSchrod Eqs}
 \hspace{1cm}
\mathcal{H}_{\ell}^{\pm}(r_*) \phi^{\pm}_{\ell,n}   =\mathcal{E}^{\pm}_{\ell,n}\phi^{\pm}_{\ell,n},\qquad n\geq 0,
\ea

With $\mathcal{W}_{\ell}$ given in Eq. (\ref{SuperPotentoial}), one can easily show that the partner potentials $U^\pm_\ell$ are related to the original axial and polar potentials via $U^\pm_\ell= V^\pm- \omega_{\mathrm{sp}}^2$.  Correspondingly, the Schrodinger-like equation \eqref{SUSYSchrod Eqs} is equivalent to the original Eq. \eqref{master} with $\big(\mathcal{E}^{\pm}_{\ell, n}\big)^2=\big(\omega^{\pm}_{\ell ,n}\big)^2 -  \omega_{\mathrm{sp}}^2 $. Note that $\omega^{\pm}_{\ell ,n}$ represents the spectrum of polar and axial perturbations at the level $n$.
Furthermore, the values of the superpotential at the endpoints of its domain
have opposite signs, $\mathcal{W}_{\ell}(0)=+\infty$, and $\mathcal{W}_{\ell}(1)<0$, so the SUSY is unbroken \cite{Gangopadhyaya:2017wpf}.

Due to the forms of $\hat{\mathcal{D}}_{\ell}^{\pm}$ and their relations to
$\mathcal{H}_{\ell}^{\pm}$, it can be shown that $\mathcal{E}^{\pm}_{\ell, n}\geq 0$. Furthermore,  if $\hat{\mathcal{D}}_{\ell}^{\pm} \phi^{\pm}_{\ell, 0}=0$ then $\mathcal{H}_{\ell}^{\pm} \phi^{\pm}_{\ell, 0}=0$, which means $\phi^{\pm}_{\ell, 0}$ is an eigenvector of $\mathcal{H}_{\ell}^{\pm}$ with eigenvalue zero, $\mathcal{E}^{\pm}_{\ell, 0}=0$ (ground state).
Correspondingly, the ground state eigenfunction can be written as $\phi^{\pm}_{\ell,  0}(r_*)= N^{\pm} \,e^{\mp\int_{x_{0}}^{r_*}\mathcal{W}_{\ell}(x) \dd x}$, where $N^{\pm}$ are normalization constants.
From $\mathcal{W}_{\ell}$ given in Eq. (\ref{SuperPotentoial}),  $\phi^{+}_{\ell,0}$ is obtained to be as in Eq. (\ref{Z-sp}) for the ASM solution.
While $\phi^{+}_{\ell,0}(r)$ is regular and normalizable, in the domain of the superpotential,  the other ground state $\phi^{-}_{\ell,0}\sim \big(\phi^{+}_{\ell,0}\big)^{-1}$ is singular at $r=0, 1$ and non-normalizable; this confirms that the SUSY is unbroken.


Since $\mathcal{H}_{\ell}^{\pm}$ are SUSY partners, their eigenfunctions and eigenvalues for $n=1,2,\dots$, are related to each other by,
\ba
\label{SUSYTrans1}
&\phi^{\mp}_{\ell,n}(r_{*})=\dfrac{1}{\sqrt{\mathcal{E}^{\pm}_{\ell\,n\pm1}}} \hat{\mathcal{D}}^{\pm}_{\ell}(r_{*})\,\phi^{\pm}_{\ell,n\pm1}(r_{*}),\\
\label{Isospect}
 & \mathcal{E}^{-}_{\ell\,n-1}=\mathcal{E}^{+}_{\ell\,n}.
\ea
Here the level number  $n$ for $\phi^{-}_{\ell,n}$ is redefined such that the lowest normalizable sate of $ \mathcal{H}_{\ell}^{-}$ is $\phi^{-}_{\ell,0}$.

As mentioned,   the Schrodinger-like equation \eqref{SUSYSchrod Eqs} is equivalent to the original equation \eqref{master}.  Consequently, the eigenfunctions of polar perturbations are given by $Z^{+}_{\ell, n}=\phi^{+}_{\ell,n}$. In particular,  the ground state is given by  $Z^{+}_{\ell,n=0}(r)=  \phi^{+}_{\ell,0}(r)$,  with $\phi^{+}_{\ell,0}(r)$  as given in Eq. (\ref{Z-sp}) and with the spectrum $\omega^{+}_{\ell,n=0}=\omega_{\mathrm{sp}}$.  Furthermore, the eigenfunctions of axial perturbations are given by $Z^{-}_{\ell,n}=\phi^{-}_{\ell,n}$. Finally, from Eqs. \eqref{SUSYTrans1} and \eqref{Isospect} and the fact that there are only $\ell-2$ axial bound states, we obtain, 
 \ba
 \label{RW-Z Transform1}
&Z^{\mp}_{\ell,n}(r_{*})=\dfrac{1}{\sqrt{(\omega^{\pm}_{\ell\,n\pm1})^2-\omega^{2}_{sp(\ell)}}} \hat{\mathcal{D}}^{\pm}_{\ell}(r_{*})Z^{\pm}_{\ell,n\pm1}(r_{*}) \, , 
\ea
with $\omega^{+}_{\ell,n}=\omega^{-}_{\ell,n-1}$ for $n=1,2,\dots, \ell-2$.

The above analysis confirms that the isospectrality between polar and axial gravitational perturbations is preserved in the Schwarzschild interior. This isospectrality arises because $U^\pm_\ell$ are partner potentials in the SUSY sense, and the singularity of the potentials at $r=0$ does not destroy this property. In addition, the polar perturbations contain an extra mode that does not have any correspondence in axial modes. This is nothing but the ASM $Z^{+}_{\ell,n=0}=Z^{+}_{\ell,\mathrm{sp}}$ with the eigenfrequency
$\omega_{\mathrm{sp}}$.\\

\section{Numerical Results and Implications}

Our numerical results also confirm the isospectrality and the existence of the ASMs. We present the detailed numerical analysis in the Appendix \ref{numeric-ap}. Below is the summary of our main numerical results and their physical implications.

({\bf a}): For a given value of $\ell$, there are total $\ell-1$ bound state solutions for the polar perturbations. This should be compared with the case of axial perturbations, which admit $\ell-2$ solutions \cite{Firouzjahi:2025jja}. ({\bf b}): All $\ell-2$ spectra of the axial perturbations are identically shared with the polar perturbations. In this view, the isospectrality between the polar and axial perturbations does hold. We have confirmed this conclusion to 40 digits in our numerical analysis. ({\bf c}): The additional mode in polar perturbations is indeed the ASM with the spectrum given in
Eq. (\ref{special}). This is the ground state solution for the polar perturbations in which, for a given value of $\ell$,
$\omega^2$ has the least value (i.e., the most negative value). In particular, for $\ell=2$, $\omega_{\mathrm{sp}}$ is the only allowed spectrum. We have confirmed that the spectrum of ASM agrees
with Eq. (\ref{special}) to 40 digits. This is significant, noting that there are controversies on the exact numerical values of ASMs for QNMs \cite{Berti:2009kk, MaassenvandenBrink:2000iwh, Leaver:1985ax, Nollert:1993zz, Liu:1996cxr}. ({\bf d}): The level number of the excited states with $n=1, 2,..., \ell-2$ is the same as the number of the nodes in the profile of $Z(r)$. For example, the first excited state has one node, the second excited state has two nodes, and so on. On the other hand, the most excited state with $n=\ell-2$ has $\ell-2$ nodes. ({\bf e}): For large $\ell$ and for low-lying states (say $n= 1$), the spectrum satisfies an analytic Wentzel-Kramers-Brillouin (WKB) formula as given in \cite{Firouzjahi:2025jja} in which $\omega_I \propto (\ell+ \frac{1}{2})^4$. ({\bf f}): For $\ell \rightarrow \infty$, the spectrum of most excited states with $n=\ell-2, \ell-3,...$, are equally  spaced with $\omega_I = (\ell-n)-1$.  An important conclusion from the above results is that independent of $\ell$ and $n$, all modes satisfy the lower bound $\omega_I >1$ with the most excited state having the minimum value of $\omega_I$ \cite{Firouzjahi:2025jja}. Note that the conclusions ({\bf d}), ({\bf e}), and ({\bf f}) listed above are identical to axial perturbations obtained in \cite{Firouzjahi:2025jja}. Furthermore, conclusions  ({\bf b}) and ({\bf c}) are confirmed analytically from our SUSY analysis as reviewed in the previous section.

\begin{table*}
\centering
\scalebox{1.2}{
\begin{tabular}{|c||c|c|c|c|c|c|}
\hline
$n$ & $\ell=2$ & $\ell=3$ & $\ell=4$ & $\ell=5$ & $\ell=6$ & $\ell=7$\tabularnewline
\hline
0 & 4.000000000 & 20.00000000 & 60.00000000 & 140.0000000 & 280.0000000 & 504.0000000\tabularnewline
\hline
1 & $-$ & 1.705649856 & 5.745972088 & 13.76767450 & 27.78537685 & 50.20490178\tabularnewline
\hline
2 & $-$ & $-$ & 1.394461407 & 3.968693503 & 8.389356734 & 15.42616838\tabularnewline
\hline
3 & $-$ & $-$ & $-$ & 1.274578721 & 3.315604823 & 6.487859859\tabularnewline
\hline
4 & $-$ & $-$ & $-$ & $-$ & 1.211013017 & 2.982338127\tabularnewline
\hline
5 & $-$ & $-$ & $-$ & $-$ & $-$ & 1.171550415\tabularnewline
\hline
\end{tabular}
}
\vspace{0.5cm}
\caption{$\omega_I$ for polar perturbations for different values of $\ell$. For each $\ell$, there are $\ell - 1$ bound states in which the ASM is the ground state with the largest value of $\omega_I$ (most negative value of $\omega^2$) while the most excited state with $n=\ell-2$ has the lowest value of $\omega_I$. Here and in Fig. \ref{fig:Z},  $\omega_I$ are normalized in units of $(2GM)^{-1}$.}
\label{table:omegaI}
\vspace{.5cm}
\end{table*}


\begin{figure*}
\begin{minipage}[b]{1\textwidth}
\subfigure[\label{fig:Z2}]{\includegraphics[width=0.48\textwidth]
{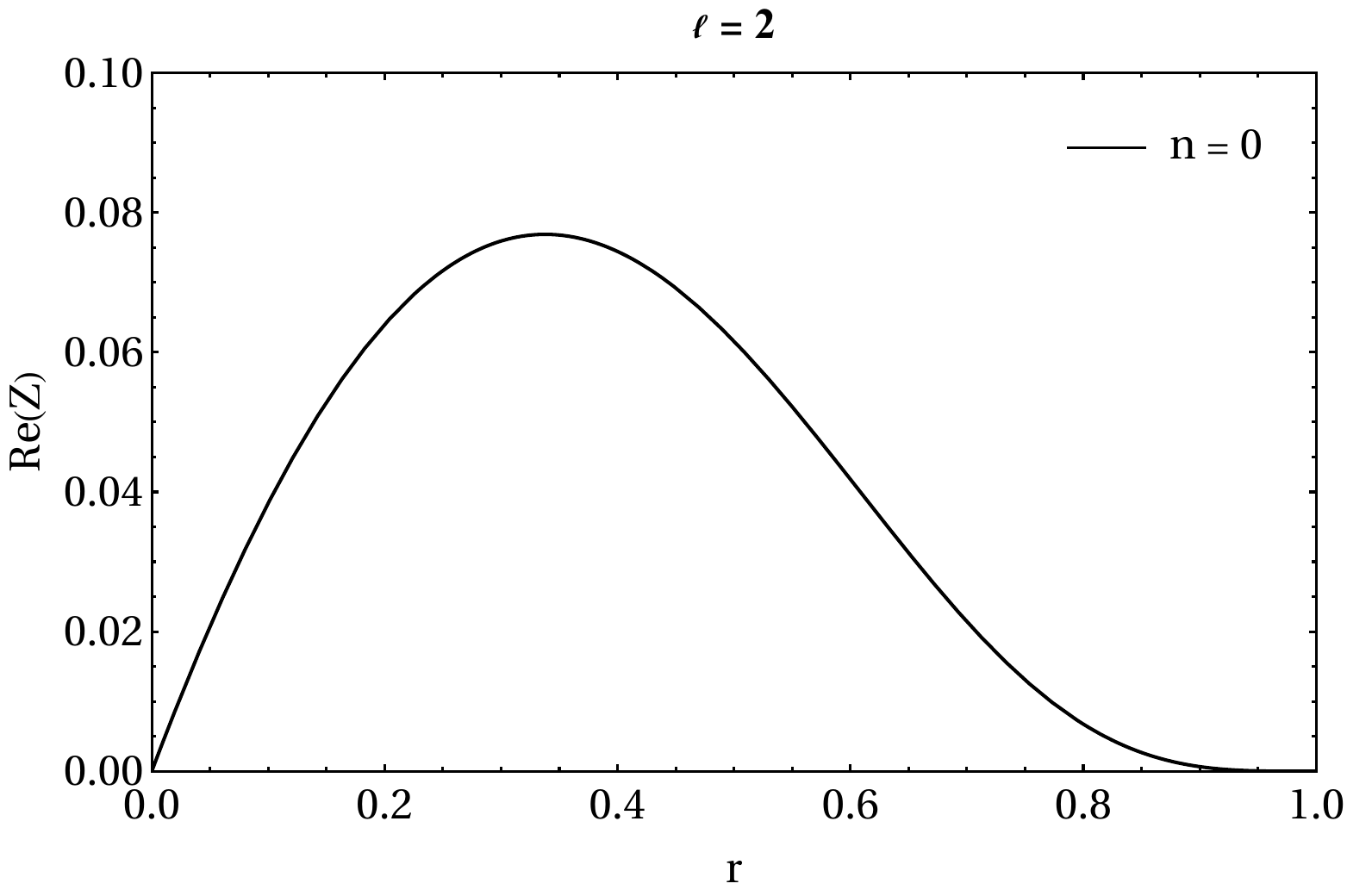}}\hspace{.1cm}
\subfigure[\label{fig:Z3}]{\includegraphics[width=0.48\textwidth]
{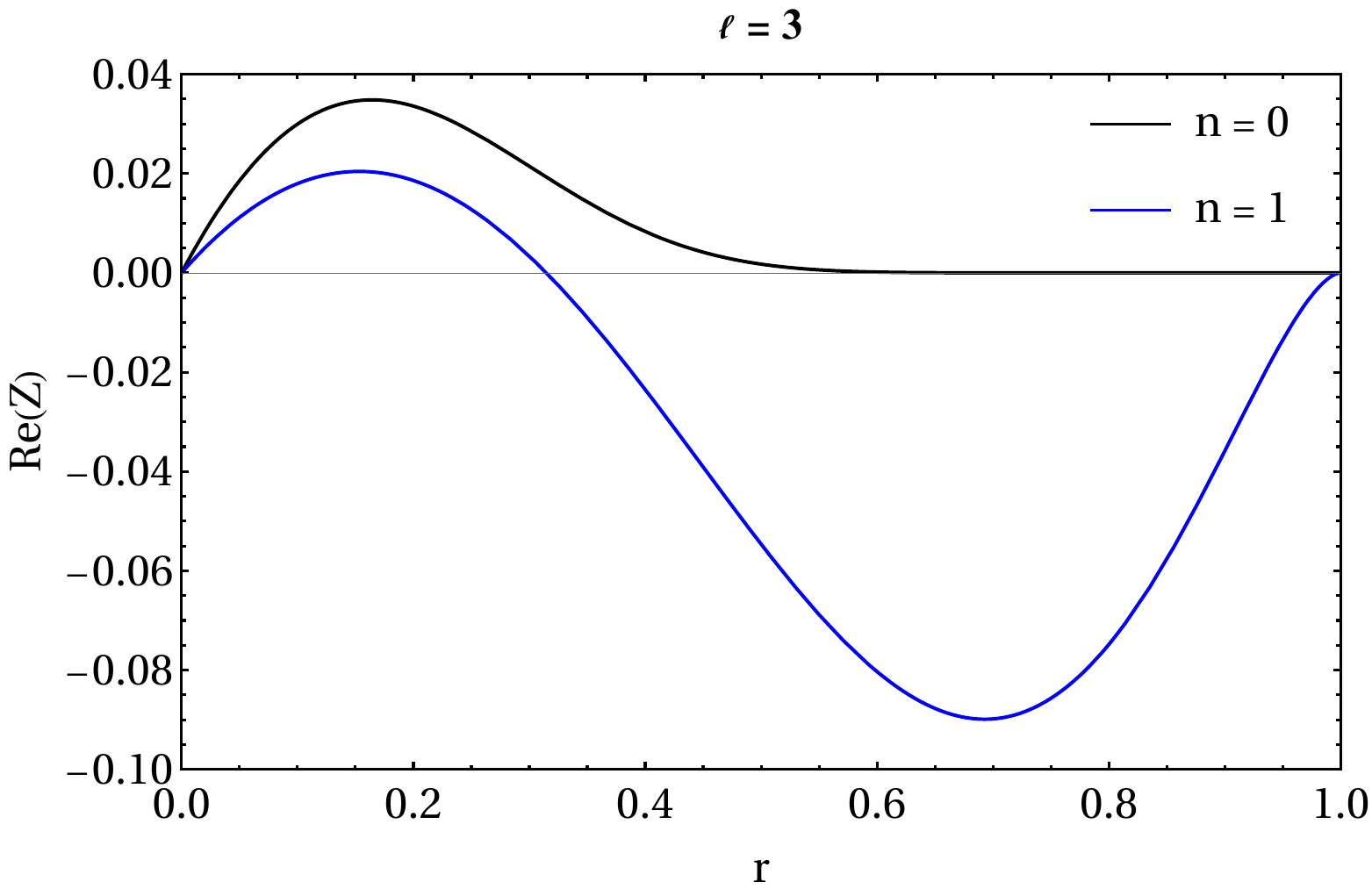}}
\subfigure[\label{fig:Z4}]{\includegraphics[width=0.48\textwidth]
{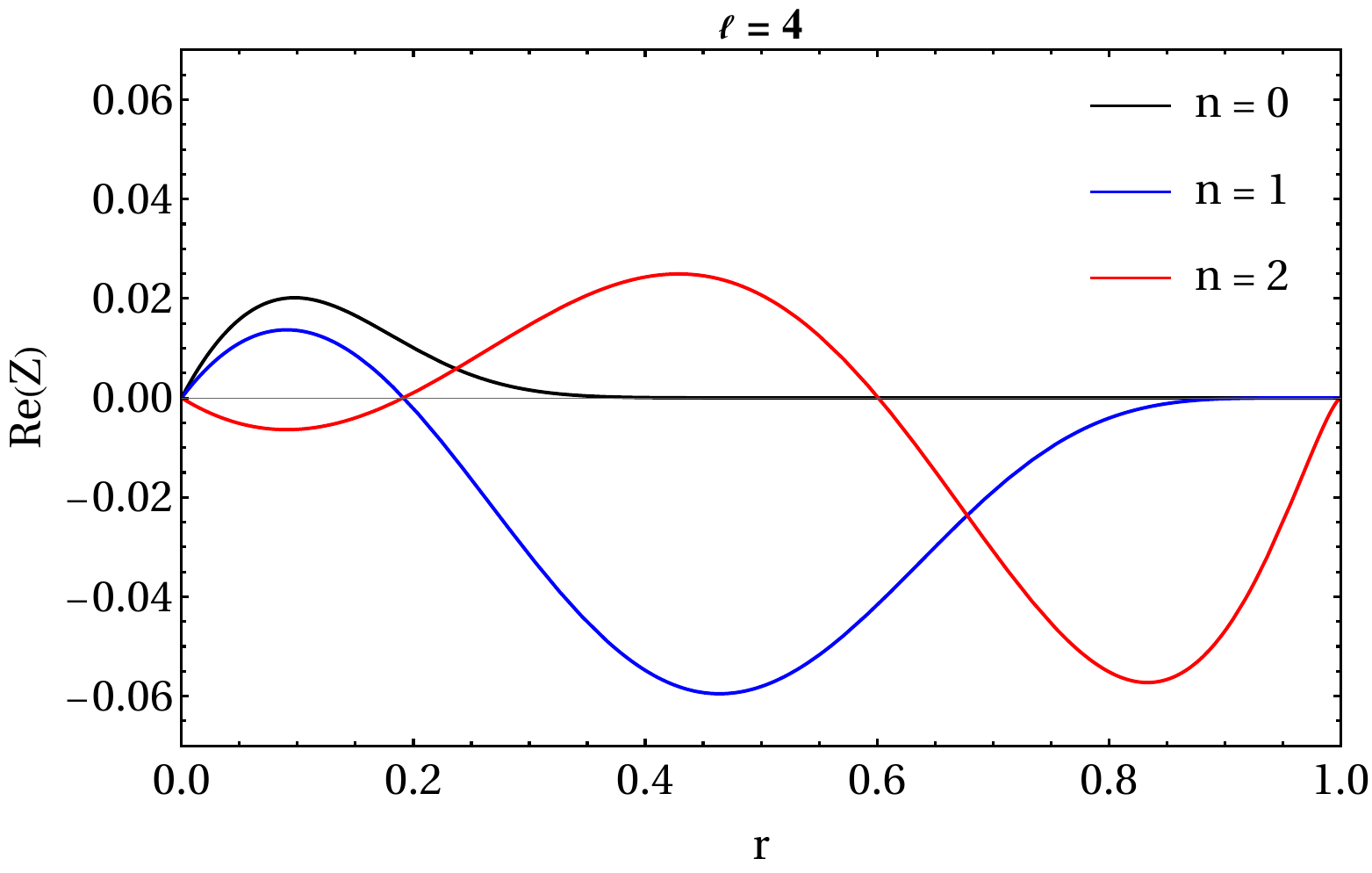}}
\subfigure[\label{fig:Z5}]{\includegraphics[width=0.48\textwidth]
{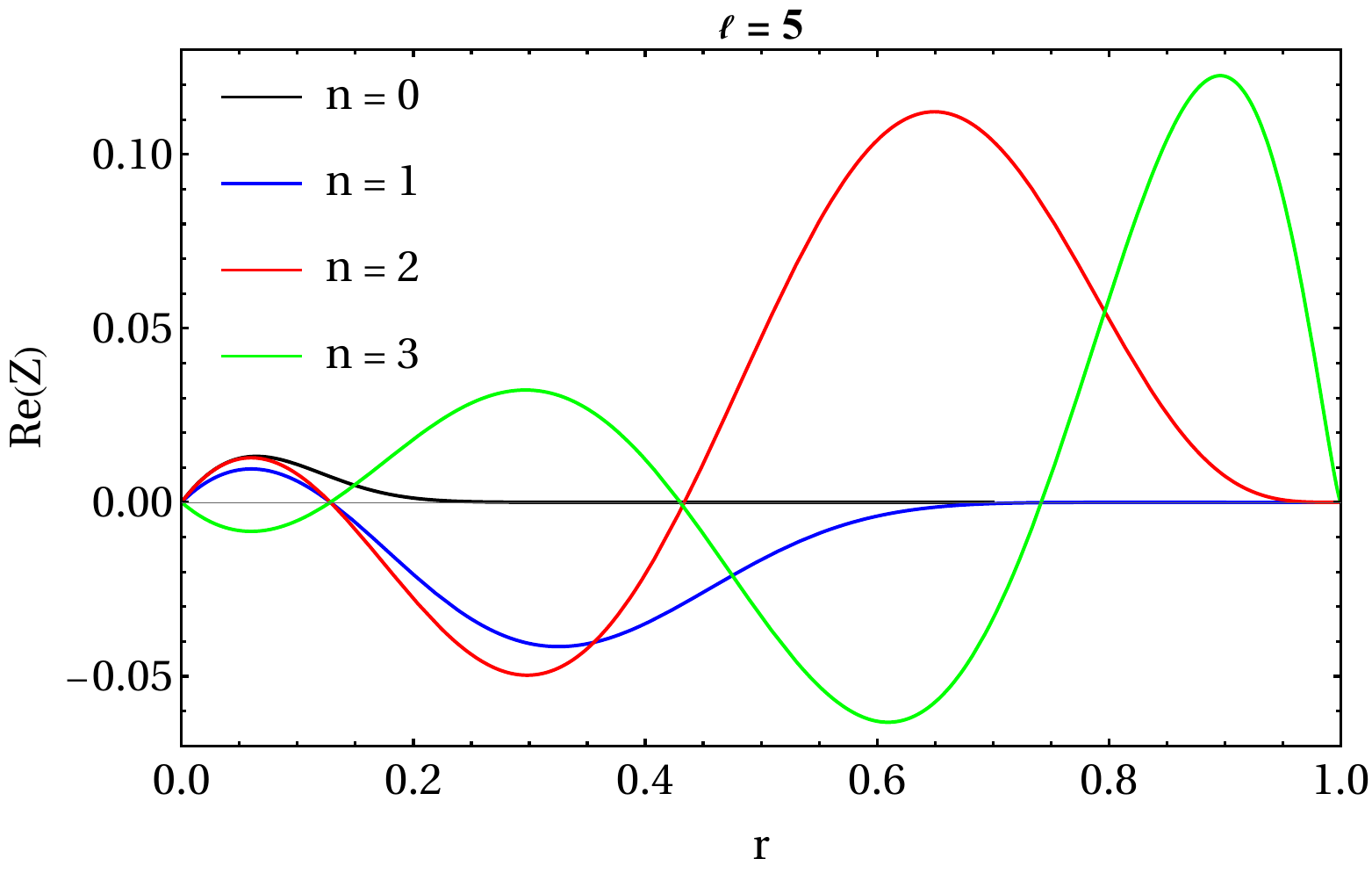}}
\end{minipage}
\caption{$\Re(Z^+)$ for polar perturbations for some values of $\ell$. For a given $\ell$, there are $\ell-1$ bound states with $n=0,1,,..., \ell-2$. Each solution with state level $n$ has $n$ nodes. The ASM is the ground state with no node. For a better visualization, unlike  Fig. \ref{special-fig},
$Z^+$ is not normalized here. Panels (a)–(d) correspond to $\ell = 2,3,4$, and $5$, respectively.}
\label{fig:Z}
\end{figure*}

In Table \ref{table:omegaI}, we have presented $\omega_I$ for some values of $\ell$. We have calculated the spectrum in 40 digits, but in order to save space, here we only present the spectrum to 10 digits. One can check that, except for $\omega_{\mathrm{sp}}$ associated to the ground state $n=0$, all the remaining $\omega_I$ are the same as in axial perturbations presented in \cite{Firouzjahi:2025jja}.

In Fig. \ref{fig:Z}, we have presented the plots of $\Re(Z)$ for the polar perturbations for some values of $\ell$ and for all values of $n=0,1,..., \ell-2$. To obtain these plots, we have solved the differential equation \eqref{d2g} numerically. In these figures, we see that the profile of the ground state with $n=0$ has no node while the profile of the excited states with $n\ge 1$ has exactly $n$ nodes.

As mentioned in point ({\bf f}) listed above, the spectrum of perturbations for the most excited states $n=\ell-2, \ell-3,...$
have the important universal property that $\omega_I$ are equally spaced for $\ell \rightarrow \infty$. This was shown for $s=0, 1, 2$ in \cite{Firouzjahi:2025jja}, and following the same analysis, one can check that this is true for $s=1/2, 3/2$ as well.  This has important implications for the understanding of the BH area quantization as follows.

Following the logic of \cite{Hod:1998vk, Maggiore:2007nq}, one can use  Bohr's correspondence principle for the transition between two states of $n, n' \gg1$  to obtain a semi-classical expression for Bekenstein's BH area quantization \cite{Bekenstein:1974jk}.  For this prescription to be applicable, the spectrum should depend only on the BH intrinsic properties and not on the properties of the fields under consideration. Indeed,  the bound states in the interior of BHs satisfy this universal property in which,  independent of $s$,  $(2 G M)\, \omega_I (n) \rightarrow (\ell-n)-1 $ for $n\rightarrow \infty$ (here we have restored the scaling $2 G M$).
Now consider the transition from the most excited state $n=\ell-2$ to the second most excited state $n'= \ell-3$. This yields $( 2 G M) \Delta \omega = 1$, inducing  a change in the mass of BH, $\Delta M= \hbar \Delta \omega=
\hbar ( 2 G M)^{-1}$. On the other hand, from the BH area formula
$A= 16 \pi G^2 M^2$ we have $\Delta A= 32 \pi G^2 M \Delta M$. Using our predictions for $\Delta M$, this yields the BH area quantization,
\ba
\hspace{0.5 cm} \Delta A= 32 \pi G^2  M \Delta M= 16 \pi l_{\mathrm{Pl}}^2 \, ,   \quad \quad (l_{\mathrm{Pl}}^2 \equiv \hbar G) \, .
\ea
In comparison, Maggiore \cite{Maggiore:2007nq} has used  the transition between highly excited states of QNMs in the exterior region and  obtained
$\Delta A = 8 \pi l_{\mathrm{Pl}}^2$.
Indeed, parameterizing the BH area quantization via $\Delta A = \alpha l_{\mathrm{Pl}}^2$, there are various estimations  for the numerical factor $\alpha$ ranging from $4 \ln 2$ to $32 \pi$ \cite{Bekenstein:1974jk, Bekenstein:1995ju, Mukhanov:1986me, Hod:1998vk, Maggiore:2007nq, Coates:2021dlg} while our prediction is $\alpha= 16 \pi$.

Finally, it is helpful to examine the similarities and differences of our bound states with QNMs in the exterior region. In both cases, the isospectrality is valid, and the polar perturbations have the additional ASM excitations. However, a key difference is that for a given value of $\ell$, the number of bound states in the interior of the BH  is finite, but the number of QNMs is infinite. This is due to differences between the asymptotic shapes of the corresponding effective potentials in the interior and exterior regions. More specifically, $V^-_{\mathrm{eff}}$ can be approximated by the Coulomb potential in far regions \cite{Liu:1996cxr}. However,  near the horizon in the interior $(r_* \rightarrow -\infty)$ it falls off exponentially like $e^{r_*}$ \cite{Firouzjahi:2018drr} (see Appendix \ref{numeric-ap}).It is understood in the literature that the potentials with exponential fall off have a finite countable spectrum.  This is similar to the Morse potential in which the potential falls off exponentially and the number of bound states is finite, see~\cite{Bargmann} and \cite{Gangopadhyaya:2017wpf}. More specifically, the finiteness of the number of bound states originates from the asymptotic form of the potential for $r_* \rightarrow -\infty$ wherein the domain of potential is a semi-infinite interval. The Bargmann limit  \cite{Bargmann} gives the following upper value for the number of bound states ($N_\ell$) for the central potential $V(r)$,
\begin{equation}
N_{\ell}<\dfrac{1}{2\ell+1}\int_{0}^{+\infty}r|V(r)|\,dr ,
\end{equation}
see also the Schwinger analysis \cite{Schwinger}. One can extend the Bargmann-Schwinger analysis for the RW potential in the Schwarzschild BH, demonstrating the finiteness of the number of bound states for the interior region.

\vspace{0.3cm}

\section{Conclusions}

In this work, we have studied the bound states of the polar perturbations in the interior of the Schwarzschild  BH and examined various of its physical implications.
The bound states are solutions with an imaginary spectrum
which are regular at $r=0$  while their profiles fall off exponentially near the event horizon.

We have shown numerically that for a given value of $\ell$, there are $\ell-1$ bound states of polar perturbations in which the spectra of $\ell-2$ of these bound states are identical (within 40 digits in our numerical analysis) with those of axial perturbations. As a result, the isospectrality between these two perturbations does hold. However, there is an additional mode in the spectrum of polar perturbations which turns out to be the ASM with the spectrum Eq. (\ref{special}) and the mode function  Eq. (\ref{Z-sp}). The isospectrality between the polar and axial perturbations is a direct consequence of the map that links these two perturbations, as we studied via the SUSY method. The fact that the potentials are singular at $r=0$ does not destroy this correspondence.  Unlike the QNMs spectrum, the number of bound states is finite (for a given value of $\ell)$,  with ASM being the ground state. In Table \ref{table:omegaI}, we have presented the numerical results for $\omega_I$ for some values of $\ell$, while the profiles of the mode functions are presented in Figs. \ref{special-fig} and \ref{fig:Z}. From Fig. \ref{fig:Z}, we observe that the number of nodes is equal to the level number  $n$. As in \cite{Firouzjahi:2025jja}, and
independent of $s$, the bound states associated with the highly excited states are equally spaced. Employing this universal property, the BH area quantization in semi-classical approximation is calculated to be $\Delta A=16 \pi l_{\mathrm{Pl}}^2$.

\vspace{0.7cm}

\section*{Acknowledgment}

We thank Bahram Mashhoon and Antonio Riotto for helpful discussions and correspondences. The work of H. F. is partially supported by INSF of Iran under the grant numbers 4046375 and 4045105. \\
\vspace{0.3cm}

%


\appendix


\section{Numerical Analysis}
\label{numeric-ap}

Here we present the numerical analysis in some detail.

The goal is to solve Eq. (\ref{master}) numerically for polar perturbations, in which the axial and polar potentials are given by respectively, 
\ba
\label{V-RW}
V^{-}(r_*)= \big( 1- \frac{1}{r}\big) \Big( \frac{\ell (\ell+1)}{r^2} -\frac{3}{r^3}\Big) \, ,
\ea
and,
\ba
\label{V-Z}
V^{+}(r_*)= \frac{r-1}{r^4} \Big[ \frac{8 \lambda^2 (\lambda+1) r^3 + 12 \lambda^2 r^2 + 18 \lambda r + 9}{(3+ 2 \lambda r)^2} \Big] \, .
\ea
Note that in the interior of BH,
$r_*= r + \ln \big( 1-r \big)$.

One can check that near $r=0$, $V^- \simeq 3/r^4 \simeq 3/4r_*^2$ while $V^+\simeq -1/r^4 \simeq -1/4 r_*^2$.
On the other hand, noting that  near the horizon $r-1 \simeq - e^{r_*}$, 
one can show that,
\ba
\hspace{1cm}V^-(r_*\rightarrow -\infty)  \simeq (1-2 \lambda) e^{r_*} \, ,
\ea 
and
\ba
\hspace{1cm} V^+(r_*\rightarrow -\infty) 
\simeq  -\frac{4 \lambda^2 + 4 \lambda +3}{3 + 2 \lambda}  e^{r_*} \, .
\ea
The plots of $V^\pm(r_*)$ given in  Eqs.  \eqref{V-RW} and \eqref{V-Z}   for $\ell = 3$ are presented in Fig. \ref{fig:Veff}. 

Employing the regularity of the wave function at $r=0$ and pulling out the asymptotic form of the bound state at $r=1$, and similar to Eq. (\ref{Z-sp}) for ASM, we consider the following ansatz for the general polar perturbations,
\begin{equation}
\label{Z-g}
Z^+(r)\equiv\frac{r(r-1)^{-i\omega}e^{-i\omega r}}{\left(\lambda r+\frac{3}{2}\right)^{2}}g(r) \, ,
\end{equation}
in which the auxiliary function $g(r)$ is regular everywhere inside the BH. Note that $Z^+(r)$ is constructed from the components of polar metric perturbations \cite{Chandrasekhar:1985kt}. The regularity of metric perturbations requires that $Z^+(r) \rightarrow r$ near the center of the BH. Therefore, with the scaling defined in Eq. (\ref{Z-g}),  the function $g(r)$ should be regular at the center.

Rewriting the corresponding Zerilli equation \eqref{master} in $r$ coordinates and  plugging back the ansatz Eq. (\ref{Z-g}) in Eq. \eqref{master}, we obtain the following equation for the auxiliary function $g(r)$,
\begin{align}
& r(r-1)(2\lambda r+3)^{2}\frac{\dd^{2}g}{\dd r^{2}}-(2\lambda r+3)
\nonumber
\\
& \times\bigg[4i\omega\lambda r^{3}+(4\lambda+6i\omega)r^{2}-6(\lambda+1)r+3\bigg]\frac{\dd g}{\dd r}
\nonumber
\\
& -2\bigg[4\lambda^{3}r^{2}+4\lambda^{2}r\left(3-i\omega r^{2}\right)+3\lambda(4r-1) +9i\omega r\bigg]g=0 \, .
\label{d2g}
\end{align}

As mentioned before, the function $g(r)$ is regular in the interval $0< r<1$ so we can use the following Frobenius series expansion,
\begin{equation}
\label{g-sum}
g(r)=\sum_{m=0}^{\infty}c_{m}r^{m} \, .
\end{equation}
Substituting the above series expansion into Eq. \eqref{d2g} leads to the following five-term recursion relation $c_m, c_{m-1}, c_{m-2}, c_{m-3}, c_{m-4}$, 
\begin{align}
& c_{m}=\frac{1}{3m^{2}}\Big[(3-4\lambda)m^{2}+(16\lambda-3)m-10\lambda\Big]c_{m-1}
\nonumber
\\
& -\frac{2}{9m^{2}}\Big[2\lambda^{2}\left(m^{2}-8m+18\right)-6\lambda\left(m^{2}-5m+4\right)
\nonumber
\\
& +9i\omega(m-1)\Big]c_{m-2}+\frac{4\lambda}{9m^{2}}\Big[\lambda\left(m^{2}-9m+18\right)
\nonumber
\\
& -6i\omega(m-3)-2\lambda^{2}\Big]c_{m-3}-\frac{8i\omega\lambda^{2}(m-5)}{9m^{2}}c_{m-4} \, ,
\label{cm}
\end{align}
with the following additional relations between the coefficients $c_1, c_2, c_3$ and $c_0$,
\begin{align}
c_{1} &= \frac{2}{3}\lambda c_{0} \, ,
\label{c1}
\\
c_{2} &= -\frac{1}{6}\left(2\lambda^{2}+2\lambda+3i\omega\right)c_{0} \, ,
\label{c2}
\\
c_{3} &=-\frac{1}{9}(\lambda+1)\Big(2\lambda(\lambda+1)+3i\omega\Big)c_{0} \, .
\label{c3}
\end{align}

Our five-term recursion relation here should be compared with the corresponding three-term recursion relation for axial perturbations \cite{Leaver:1985ax, Nollert:1993zz, Firouzjahi:2025jja}. This is because the Zerilli equation is algebraically more complicated than the RW equation.

For any value of $m\geq4$, the recursion relation \eqref{cm} and the conditions ‌\eqref{c1}-\eqref{c3} give a polynomial expression for $c_m$ in terms of $\omega_I$ with a normalization coefficient proportional to $c_0$.
As supposed, the function $g(r)$ is regular at $r=0$, and in order for this function to be regular at $r=1$ too, it is necessary that the sum $\sum_{m=0}^{\infty}c_{m}$ exists and is finite. This criterion determines the spectrum of our bound state perturbations. Note that we are looking for the bound state solution with an imaginary spectrum $\omega=i\omega_{I}$ with $\omega_I > 0$.

We use the numerical method prescribed in \cite{Firouzjahi:2018drr, Firouzjahi:2025jja} to calculate the bounded-state spectrum. As argued in \cite{Firouzjahi:2018drr}, in order for the sum $\sum_{m=0}^{\infty}c_{m}$ to exist and to be finite, it is sufficient that the coefficient $c_m$ becomes zero for a large value of $m$. If this condition holds for a large value of $m$, then the successive terms $c_{m+1}$, $c_{m+2}$, ... are suppressed and we can expect that the sum $\sum_{m=0}^{\infty}c_{m}$  to converge quickly. The high accuracy of this method compared to the method of Leaver \cite{Leaver:1985ax} was confirmed in \cite{Firouzjahi:2018drr}. Furthermore, our smooth numerical plots for the profile of $Z(r)$ as presented in Fig. \ref{fig:Z} confirm that this approach is reliable. In addition, the fact that our numerical method predicts the existence of the ASM as a ground state solution, which is confirmed analytically as well, indicates that our numerical prescription is based on solid ground.

Moreover, we have examined the validity of our numerical prescription described above by employing another independent numerical approach, which is available online as well~\cite{rezazadeh_bound_2025}. This code works almost identically to V{\"o}lkel's method~\cite{Volkel:2025lhe}, which is based on the computation of the Wronskian $W(r)=Z_{\rightarrow}^{+}\frac{dZ_{\leftarrow}^{+}}{dr}-\frac{dZ_{\rightarrow}^{+}}{dr}Z_{\leftarrow}^{+}$, where $Z_{\rightarrow}^{+}$ and $Z_{\leftarrow}^{+}$ are the solutions of the Zerilli differential equation (in $r$ coordinate) by integrating from the right and left endpoints of the interval $[0, 1]$ respectively. To calculate these functions, we applied the eighth-order Runge-Kutta (RK8) method in our code to solve the differential equation~\eqref{d2g} by using the series~\eqref{g-sum} as boundary conditions. Our code employs a random-walk minimization method to minimize $|W|$ evaluated within the integration interval. The code is written on the basis of the message passing interface (MPI) method, which allows parallel processing and speeds up the minimization process significantly. Starting from a guess value for $\omega$ that is within the radius of convergence of the desired root, we calculate the value of $|W|$. Then, we consider a value close to the previous frequency and calculate $|W|$ for it as well. This ensures that the Wronskian value is re-evaluated at every adjustment of the eigenvalue search. If the result for the new frequency is less than the corresponding value for the previous frequency, we search around the new frequency; otherwise, we return to the previous one. We repeat this procedure until we reach the desired accuracy for the bound state frequency. We have checked that the results obtained from this independent method confirm our numerical results with very good accuracy.


\bibliography{polar}





\end{document}